**Comparative evaluation of different methods of "Homomorphic Encryption" and "Traditional Encryption" on a dataset with current problems and developments**

Tanvi S. Patel[1†], Srinivasakranthikiran Kolachina[1†], Daxesh P. Patel[2], Pranav S. Shrivastav[2*]

[1]Department of Computer Science, Bowie State University, Bowie, MD 20715-9465, USA

[2]Department of Chemistry, School of Sciences, Gujarat University, Navrangpura, Ahmedabad 380009, Gujarat, India.

† TSP and SK contributed equally to this study

**\* Correspondence to:**

Pranav S. Shrivastav,

Department of Chemistry, School of Sciences, Gujarat University,

Navrangpura, Ahmedabad 380009, Gujarat, India.

E-mail: pranav_shrivastav@yahoo.com

**Abstract**

A database is a prime target for cyber-attacks as it contains confidential, sensitive, or protected information. With the increasing sophistication of the internet and dependencies on internet data transmission, it has become vital to be aware of various encryption technologies and trends. It can assist in safeguarding private information and sensitive data, as well as improve the security of client-server communication. Database encryption is a procedure that employs an algorithm to convert data contained in a database into "cipher text," which is incomprehensible until decoded. Homomorphic encryption technology, which works with encrypted data, can be utilized in both symmetric and asymmetric systems. In this paper, we evaluated homomorphic encryption techniques based on recent highly cited articles, as well as compared all database encryption problems and developments since 2018. The benefits and drawbacks of homomorphic approaches were examined over classic encryption methods including Transparent Database Encryption, Column Level Encryption, Field Level Encryption, File System Level Encryption, and Encrypting File System Encryption in this review. Additionally, popular databases that provide encryption services to their customers to protect their data are also examined.

## 1. Introduction

The security of data available in files, databases, accounts, and networks is of prime concern in the digital world. Data security is essential to evaluate threat to data and reduce the risk associated with data storage and handling. The effective data security system takes adequate measures to identify the importance of specific types of datasets and then apply the most relevant security controls. Further, to minimize data hacking, it is essential to have database encryption. Database encryption helps to protect private information, sensitive data, and can enhance the security of communication between client apps and servers. When the data is encrypted, even if an unauthorized person or entity gains access to it, it is not possible to read the data. Database encryption is a procedure that employs an algorithm to convert data contained in a database into "Cipher text," which is incomprehensible unless decrypted first. [1, 2].

Popular database encryption methods include, Always Encrypted (AE), Transparent Data Encryption (TDE) & Cell Level Encryption (CLE), Dynamic Data Masking (DDE), Vormetric Transparent Encryption (VTE), Column Level Encryption (CLE), Field Level Encryption (FLE), and Encrypting File System (EFS). These encryption methods are widely used at various databases like Azure, AWS, MySQL, Oracle, SQL, Cassandra, Postgre SQL, MongoDB and many others. **Figure-1** shows the most popular databases used worldwide. All these databases employ different encryption methods. **Table-1** shows the results from database paper survey from reported studies [3-10]. It is evident that maintaining the confidentiality of the data is one of the primary reasons using database encryption. While most of the databases use TDE encryption. SQL Server introduced Transparent Data Encryption (TDE) in 2008, which is also available in Oracle database management systems. It is a type of encryption that safeguards the database's most sensitive information. And protects the data by encrypting the database's underlying files rather than the data itself. It protects the data from being hacked and copied to another server and to open the files original encryption certificate and master key is essential. TDE is now available to encrypt data on page level, column level and file level and does real-time I/O encryption and decryption of data and log files [10]. The encryption uses a database encryption key (DEK) and maintains database boot record, which is the key for data availability during recovery. It works like a fob key for vehicles, while the access to the database lies solely with the owner. Thus, TDE is essentially based on Advanced Encryption Standard (AES), or triple Database Encryption Standard (DES).

In 1970, IBM introduced "Crypto group" to protect their customer data, which was accepted by the US as DES in 1973. It was the first encryption algorithm to be developed but was soon replaced with AES, which was founded by the National Institute of Standards and Technology (NIST) in 2000. In the meantime, researchers at MIT introduced the asymmetric encryption algorithm and named it Rivest, Shamir, and Adleman (RSA). Since then, it has been one of the most widely used database encryption tools.

After the turn of the century, the business began to rely on cloud-based databases. Nowadays, with the increase in the size of databases, more and more data are stored on the cloud and thus is in high demand. However, with greater data transformation happening through the clouds, it is more prone to hacking and cybercrime. To overcome this, Craig Gentry in 2009 gave the concept of homomorphic encryption which was used for privacy-preserving outsourced storage and computation. This allowed data to be encrypted and outsourced to commercial cloud environments for processing, while being encrypted. In the present study we have reviewed the work done on homomorphic encryption and assessed types of databases that can be encrypted using homomorphic encryption.

## 2. Background on homomorphic encryption and literature review

Since the introduction of homomorphic encryption, it has remained an open problem in cryptography. Homomorphic encryption is a type of encryption that allows users to conduct specified algebraic operations directly on the ciphertext and then decrypt the result. The outcome can be compared with the result obtained by performing the same method directly on plaintext (unencrypted data). However, the advantage of homomorphic encryption is that even when data is encrypted, users may still analyze and retrieve specific encrypted data, enhance data processing efficiency, and ensure data transmission security. In the end, correct encrypted data can nevertheless yield correct decryption results [11-15]. The homomorphic techniques are classified into three categories based on the number of operations executed on encrypted data (**Figure 2**). This includes, Partially Homomorphic Encryption (PHE), which allows unrestricted single operation, Somewhat Homomorphic Encryption (SHE) allows some of the operations for limited number of times; and Fully Homomorphic Encryption (FHE) allows all types of operations to be carried out with no restriction. The practical applicability of homomorphic encryption can be achieved only if one is able to perform an arbitrary number of homomorphic additions and multiplications on encrypted data. The FHE scheme allows arbitrary operations on encrypted data, but its applicability in real applications is still a subject of further research [11-15].

In 2009, an IBM researcher proposed FHE, a new homomorphic encryption scheme based on the ideal lattice that can perform an unlimited number of operations on encrypted data that can be performed on the plaintext without decryption, allowing in-depth and infinite analysis of encrypted data without compromising its confidentiality [16]. Based on Gentry's findings, van Dijk, Gentry, Halevi, and Vaikuntanathan (DGHV) [17] developed an integer based FHE system. Even though DGHV is more concise than Gentry's system its practicality remains low. The FHE scheme proposed by Gentry [16] was further reduced by Smart [18] and Gentry and Halevi [19] in the same year, decreasing the length of the ciphertext and key, making the execution of the scheme easier. Regev [20] proposed the FHE scheme, which is by far the most concise FHE scheme, based on the Learning with Error (LWE) problem. Since then, cryptographers have developed several FHE algorithms to increase their efficiency [19, 21]

Because of its basic architecture, PHE has the most applications due to its mature scheme. It is not ideal for scenarios involving complicated query services because its function is relatively simple and can only conduct one operation. FHE is the best encryption system since it can handle complicated query requests and can theoretically be used in any situation. However, due to the high computer resource requirements, there has yet to be a solution that can be installed and run.

SHE is the result of the coordination of PHE and FHE, which can conduct a limited number of operations based on realistic computing resource requirements. Researchers have employed PHE technology and the most recent FHE technology to database privacy protection thanks to the ongoing development of homomorphic encryption technology. It will also contribute to the field's future progress and development. The application methods and techniques of homomorphic encryption in databases were summarized in the subsequent sections.

Homomorphic encryption has been studied extensively for a wide range of applications. For example, Fahsi et al. [22] presented a framework for homomorphic encryption and private information retrieval procedures in the cloud to safeguard users from illegal data access. Other homomorphic encryption algorithms exist, but they propose an upgrade for securing individuals online or in the cloud. Before focusing on current developments and breakthroughs over traditional methods, it's critical to understand the fundamentals of homomorphic encryption. According to Rohilla [23], homomorphic encryption generates ciphertexts by specific computations that provide encrypted output but requires reverse computation techniques to obtain the plaintext form of the encrypted message. FHE is a type of ring homomorphism with structure-preserving properties. This homomorphism can be additive or multiplicative; in the former, randomly selected prime numbers are added pairwise, whereas in the latter, they are multiplied [24]. Figure 3 shows the timeline of homomorphic encryption scheme from PHE to FHE.

## 3. Homomorphic Encryption Schemes

### 3.1. Partial Homomorphic Encryption Schemes

Rivest-Shamir-Adleman (RSA) devised the first homomorphic encryption system, unpadded RSA, in 1977, based on the integer factorization problem [25-27]. It is simple to calculate the product of two prime numbers but finding prime factors of a number in polynomial time is more complex. This phenomenon is the foundation of RSA's security. Because it allows multiplication on encrypted data, RSA supports homomorphic multiplication. Unpadded RSA is a PHE system because it only permits multiplication to be done an unlimited number of times. However, random bits are added before encryption to maintain the security of plaintext, resulting in the loss of homomorphic feature [25].

The first probabilistic asymmetric-key encryption technique was proposed by Goldwasser-Micali (GM) in 1982 [28, 29]. This approach is semantically secure and is based on the intractable issue of quadratic residue modulo composite N, i.e., determining if $x$ is quadratic residue modulo

N. This approach only allows for additive homomorphic operations to be performed any number of times. If the plaintext is encrypted numerous times, this approach also yields various ciphertext, but the ciphertext is too big [25]. The next PHE technique, based on a Diffie-Hellman key exchange asymmetric key encryption algorithm, was created by ElGamal in 1985 [30]. The security of this approach is based on cyclic group problems that are linked to the difficulty of computing discrete logarithms. It only performs multiplication operation any number of times.

The Paillier cryptosystem is a probabilistic asymmetric key technique invented by Pascal Paillier in 1999 [31], like GM [26]. The problem of decisional composite residuality, or identifying nth residue classes, is the foundation of this system. It also offers homomorphic addition operations for an indefinite number of times. Boneh, Dan, Eu-Jin Goh, and Kobbi Nissim invented the Boneh-Goh-Nissim (BGN) Encryption Scheme in 2005 [32]. The security of this system is dependent on the subgroup decision problem, which is based on the pairing of elliptical curves. It can be used for both homomorphic addition and multiplication, while ciphertext is always of the same size. This is a type of homomorphic encryption that allows for numerous additions but only one multiplication [25].

*3.2. Fully Homomorphic Encryption Schemes*

In the schemes described in the previous sections, just one homomorphic operation is allowed to run indefinitely, or only one operation is allowed to run arbitrarily while another is allowed to run for a limited time. Both operations are allowed arbitrarily in fully homomorphic encryption algorithms. This section looks at some of the FHE schemes that have been proposed in the literature. The approach proposed by Gentry [33] is based on ideal lattices, which are ring subsets that maintain the feature. This system allows for both addition and multiplication, however the ciphertext gets noisy after a certain threshold. To recover appropriate ciphertext from the noise, squashing and bootstrapping techniques were introduced. Even though this is a promising step toward FHE, its implementation is challenging, and it has a high computing cost for real-world applications due to complicated mathematical concepts [25]. This method has spawned numerous optimizations and new schemes. Van Dijk et al [34] suggested another FHE approach based on Gentry's bootstrapping method. Instead of using the ideal lattice, this approach used integers and is based on the problem of the Approximate Greatest Common Divisor (AGCD). It supports homomorphic additions and multiplications, but noise increases exponentially with multiplication and linearly with addition [25].

In 2011, Brakerski and Vaikuntanathan [35] suggested a fully homomorphic technique based on Ring learning with errors (RLWE) [36]. They first proposed the SHE scheme, which minimizes the worst-case hardness of problems on ideal lattices. The strategy is turned into fully homomorphic encryption schemes utilizing Gentry's squashing and bootstrapping techniques [25]. Fan and Vercauteren [37], adapted this FHE technique from LWE to RLWE to analyze and optimize a variety of subroutines in multiplication, linearization, and bootstrapping.

Another homomorphic encryption technique based on the LWE was developed in 2013 [38]. The approximation eigenvector approach is used in this system. The homomorphic operations – addition and multiplication – are identical to matrix addition and multiplication, making it easier and faster to implement than other FHE methods. Without knowing the user's public key, this approach assesses homomorphic operations. For approximate arithmetic, Jung Hee Cheon et al. [39] presented a homomorphic encryption technique. This technique supports approximate addition and multiplication of encrypted data, as well as a rescaling procedure to manage plaintext size. The proposed technique works well with transcendental functions such as exponential, multiplicative inverse, and logistic functions, as well as approximate circuits [25].

*3.3. Somewhat Homomorphic Encryption Schemes*

As previously stated, nearly all FHE techniques are achieved by "introducing noise to the plaintext." A secret key can be used to recover the message, however, with the increase of noise in the ciphertexts, they may have an accuracy problem after evaluation. Gentry devised a method known as "bootstrapping," which involves "refreshing" the ciphertext to assure decoding accuracy [16]. Since it is referred to the encryption system as "somewhat" homomorphic before bootstrapping, which is a method for noise management, this SHE nomenclature has been used in all subsequent research. As a result, every FHE scheme should include a SHE that only supports a restricted depth of evaluation circuit, as well as suitable noise management in the cyphertext.

**4. Current Approaches**

The first focus of FHE research was on lattice-based algorithms [18, 19, 40], which required relatively large public key and ciphertext sizes. The security of lattice-based systems, as well as other FHE schemes, is dependent on hard problems related to lattices, such as the sparse subset sum problem (SSSP) or the shortest vector issue (SVP). Smart and Vercauteren [41] proposed single instruction multiple data (SIMD) approaches for these systems to perform operations in parallel and thus enhance efficiency. The theoretical research on cryptography is now concentrating on FHE based on LWE and RLWE [42-44]. Regev [20] introduced LWE, which has proved to be as difficult as the worst-case lattice problems. This problem has been expanded to work over rings [45], and the efficiency of LWE has improved because of this extension. Lauter et al. [46] replicated Brakerski and Vaikuntanathan's [21] analysis of the practicality of SHE systems as well as their implementation.

Van Dijk et al. [44] proposed integer-based schemes as a theoretically easier alternative to lattice-based schemes, and they have since been further refined to provide similar functionality compared to previous lattice-based systems. Furthermore, a proposed public key compression approach has been developed. The public key was lowered in size from over 2 GB to under 1 GB [47]. The results of these integers-based methods have been improved through a batching strategy for encryption that has been proposed combining several plaintext bits into a single ciphertext [48]. The effectiveness of using this batched technique to evaluate AES over integers is comparable with the alternate implementation using the FHE technique based on RLWE [49].

There have been various optimizations in the field of homomorphic encryption to enhance efficiency and speed since the introduction of SHE and FHE [50]. Bootstrapping, as indicated in the introduction, is a costly process that homomorphically decrypts a ciphertext using a secret key that is encrypted and buried in the public key to reduce noise in the ciphertext. There has been a lot of study into enhancing or avoiding the usage of bootstrapping, such as introducing modulus switching, which helps to reduce the noise generated when performing homomorphic operations [40].

Several recent software implementations of FHE schemes have been published and an open-source software implementation of FHE is accessible online [51]. The newest online release of Halevi and Shoup's homomorphic encryption library (HElib) includes an improved version of the FHE technique [52, 53] from that of Brakerski et al. [50]. It enhances the performance of schemes by a factor of 12, compared to earlier implementations. The homomorphic AES technique requires 3 h instead of 36 h [53].

## 5. Results

In this review we tried to differentiate homomorphic and traditional machine learning encryption methods that were developed in the last four decades. Due to data privacy requirements and General Data Protection Regulation (GDPR) and California Consumer Privacy Act (CCPA), sharing private data with third parties, such as cloud services or other corporations, is difficult.

Failure to follow these rules can result in heavy penalties and a damaged business reputation. Traditional encryption methods provide an efficient and secure way to store private data on the cloud in an encrypted form. However, to perform computations on data encrypted with these methods, businesses either need to decrypt the data on the cloud, which can lead to security problems, or download the data, decrypt it, and perform computations, which can be costly and time-consuming. Homomorphic encryption enables businesses to share private data with third parties to get computational services securely [55]. With homomorphic encryption, the cloud service or the outsourcing company has access only to the encrypted data to perform required computations. These services then return the encrypted result to the owner who can decrypt it with a private key.

Any encryption scheme's performance is assessed using three criteria: security, speed, and simplicity. First and foremost, an encryption method must be secure enough from the attacker to get any type of information with a fair number of resources. Second, its effectiveness must not interfere with the user's comfort; that is, it must be transparent to the users, as users choose usability over security. Besides, other area practitioners implement an encryption system for their applications if it is easy for them to understand. When the existing FHE systems are compared against the three criteria, it becomes clear that, while progress is being made, there is still a lot of potential for improvement in all three areas, particularly in terms of speed.

On encrypted data, FHE allows an endless number of functions. However, due to limitations in the efficiency of FHE schemes, researchers are looking for SHE schemes that may be used in real-world applications [56]. Researchers have been successful in lowering homomorphic evaluation of one AES, which is a highly complex and nontrivial function to 2 s [49, 50]. Considerable efforts have been made to improve this aspect instead of constructing a new FHE scheme. The bootstrapping process, which is required to reduce noise in the assessed ciphertexts is the key factor that raises the computing cost in FHE. It is still a work in progress to create an unbounded and noise-free FHE system that allows for limitless operations without bootstrapping.

Traditional machine learning employs a centralized strategy, which necessitates the collection of training data on a single machine or in a data center. Such data must be transferred back to the central server from a specific environment. Internet of Things (IoT) and mobile network connectivity advancements have resulted in exponential data storage at the device level. As a result, traditional methods are no longer appropriate for today's high-volume data networks. We have also analyzed scientific research articles on homomorphic encryption since 1990 (**Figure 4**), with a significant increase since 2010.

While there are several benefits associated with, homotropic encryption do have some drawbacks. The intricacy of the systems is one of the most significant disadvantages. There is no overhead involved in executing computations in partially homomorphic cryptosystems, at least for the ones described [57]. FHE, on the other hand, necessitates a far more complicated lattice-based cryptosystem. Even for fundamental functions, implementing such a cryptosystem necessitates substantially more sophisticated computations and large ciphertext volumes. When utilizing the suggested security parameters, ciphertexts of 128 MB and a public key of 128 PB are created. Even when security parameters are reduced to the point where homomorphism is no longer viable, the key size involves hundreds of gigabytes, with encryption of a single bit taking up to 30 min.

## 6. Conclusion

The privacy of data has become more important than ever essentially because of the amount of information available on the internet-centric environment. It is critical to protect consumers' accounts and assets from malevolent third parties in highly sensitive systems like online shopping and e-banking. Nonetheless, it is now a standard practice to encrypt data and exchange the keys with service providers, cloud operators, and other third parties. Control over the privacy of sensitive data is lost in this paradigm. The data belongs to the users or service providers who have the key. Besides, users' sensitive data and identification credentials are with untrustworthy providers and cloud operators even after the user's engagement with the services has ended. The use of homomorphic encryption algorithms is one potential option for preserving data privacy. It is a type of encryption system that lets any third party to work on encrypted material without having to decrypt it first. Although the concept of homomorphic encryption has been established for over 30 years, Craig Gentry proposed the first conceivable and practical FHE scheme in 2009. Since then, several other FHE schemes have come up but still they need to be greatly improved to make them more practical on all platforms, as they are too expensive in real-world applications. As a result, we examined the different homomorphic encryption programs in this paper. Starting with the fundamentals, the details of the well-known PHE and SHE, both of which are critical pillars in reaching FHE, were provided. Furthermore, the implementations of Gentry-type FHE schemes and some novel advancements were elaborated. Finally, possible research avenues and lessons gained for future researchers were outlined.

Table-1 Empirical analysis of security parameters achieved using encryption methods, based on database security and encryption survey

| Security Parameters | Frequency | Percentage | Criticality |
|---|---|---|---|
| Confidentiality | 5 | 100% | High |
| Integrity | 2 | 40% | Moderate |
| Access Control | 2 | 40% | Moderate |
| Efficiency | 2 | 40% | Moderate |
| Privacy | 3 | 60% | High |

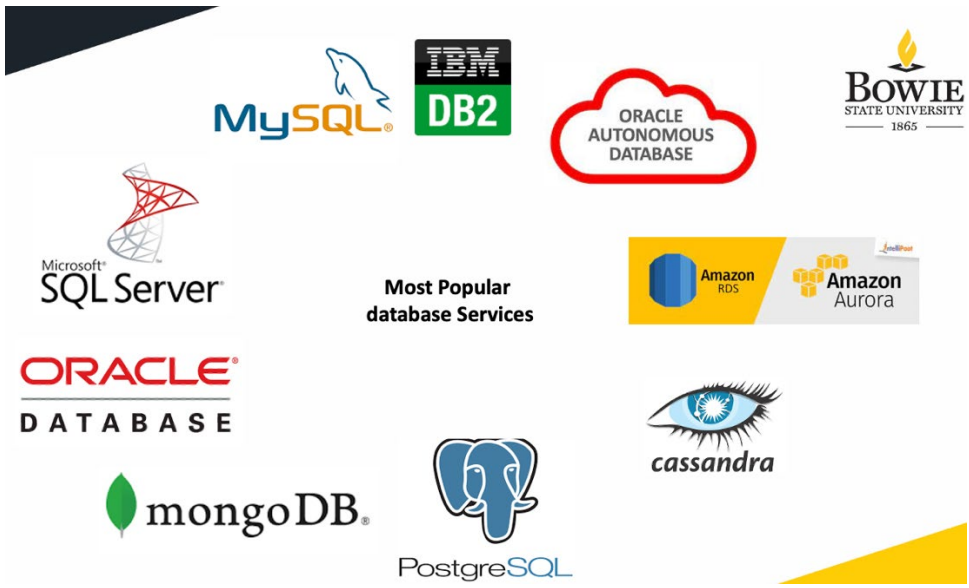

**Figure 1.** Most popular databases that are used worldwide.

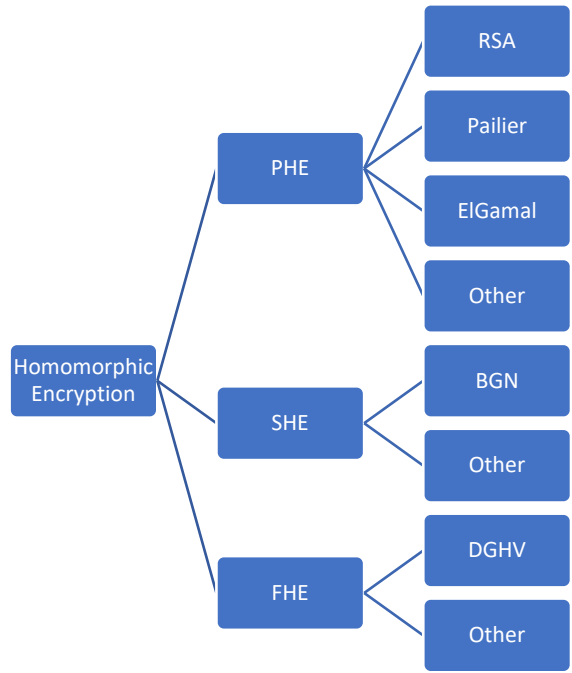

**Figure 2.** Types of homomorphic encryption

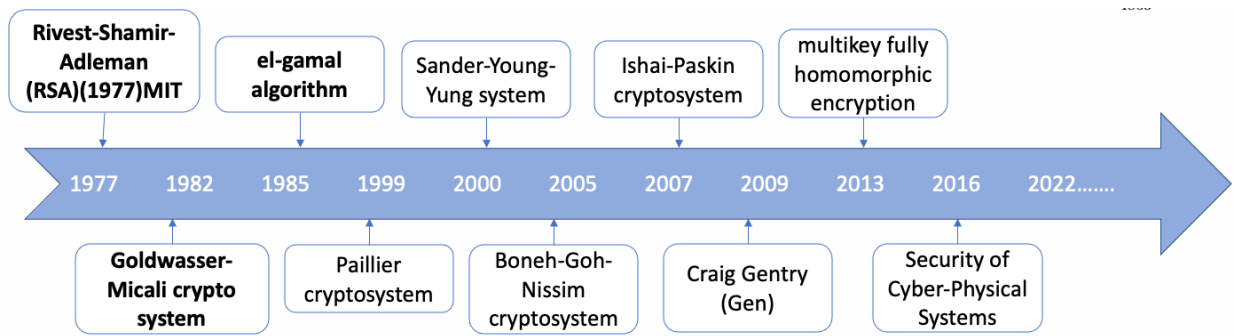

**Figure 3.** Timeline of homomorphic encryption scheme from PHE to FHE [1]

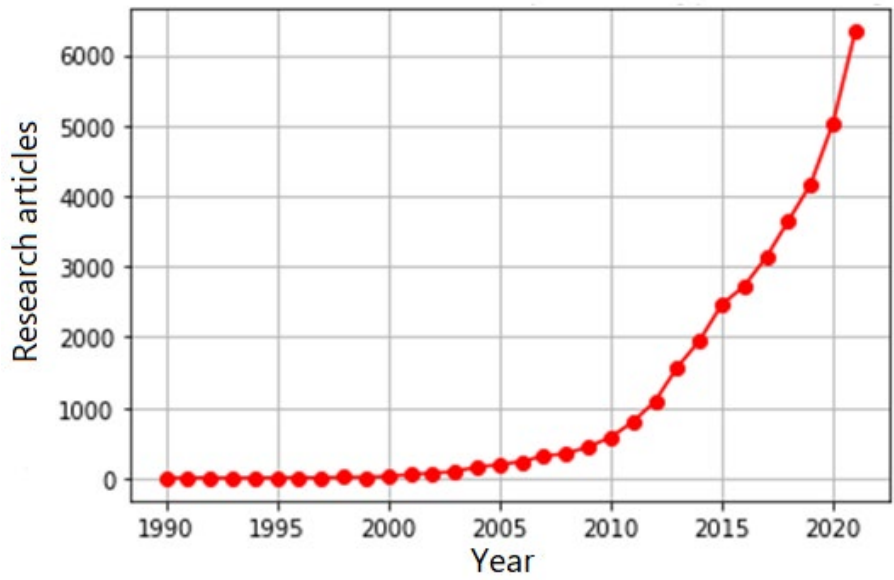

**Figure 4**. Year wise publication of research articles on homomorphic encryption